%% file: isit_2014_rev_v5.tex
\documentclass[conference]{IEEEtran}

\usepackage{amsfonts}
\usepackage{amssymb,amscd,amsthm}
\usepackage[cmex10]{amsmath}
\usepackage{mathtools} 
\usepackage{ifthen}
\usepackage[applemac]{inputenc}

\usepackage{endnotes}

\usepackage{color}

\input{./rh_ss_def.tex}

\usepackage{bm}

\usepackage{pgf,tikz}
\usepackage{pgfplots}
\usepackage{amsmath}
\usetikzlibrary{calc,shadows}
\usetikzlibrary{pgfplots.groupplots}
\usepackage{multirow}

\usepackage[applemac]{inputenc}

\usepackage{amsthm}

\pgfplotsset{width=7cm,compat=1.3}

\newtheorem{theorem}{Theorem}

\begin{document}
%
\title{Subspace clustering of dimensionality-reduced data}

\author{\IEEEauthorblockN{Reinhard Heckel, Michael Tschannen,  and Helmut B\"olcskei}
\IEEEauthorblockA{
ETH Zurich, Switzerland\\
Email: \{heckel,boelcskei\}@nari.ee.ethz.ch, michaelt@student.ethz.ch
}
}


%


\maketitle

\begin{abstract}
Subspace clustering refers to the problem of clustering unlabeled high-dimensional data points into a union of low-dimensional linear subspaces, assumed unknown. 
In practice one may have access to dimensionality-reduced observations of the data only, resulting, e.g., from ``undersampling'' due to complexity and speed constraints on the acquisition device. 
More pertinently, even if one has access to the high-dimensional data set 
 it is often desirable to first project the data points into a lower-dimensional space and to perform the clustering task there; this reduces storage requirements and computational cost.  
The purpose of this paper is to quantify the impact of dimensionality-reduction through random projection on the performance of the sparse subspace clustering (SSC) and the thresholding based subspace clustering (TSC) algorithms. 
We find that for both algorithms dimensionality reduction down to the order of the subspace dimensions is possible without incurring significant performance degradation. 
The mathematical engine behind our theorems is a result quantifying 
how the affinities between subspaces change under random dimensionality reducing projections. 
\end{abstract}


%
\IEEEpeerreviewmaketitle

\newcommand{\Ss}[1]{\mathcal{S}_{#1}}
\newcommand{\Rn}[1]{\mathbb{R}^{#1}}
\newcommand{\Us}[1]{\mathbf{U}^{(#1)}}
\newcommand{\Ph}{\mathbf{\Phi}}
\newcommand{\N}[1]{\mathcal{N}(0,#1)}
\newcommand{\mat}[1]{\mathbf{#1}}
\newcommand{\matt}[1]{\mathbf{\tilde{#1}}}

\newcommand{\Bhati}[1]{\boldsymbol{\hat{\beta}}_{#1}}
\newcommand{\bbeta}{\boldsymbol{\beta}}

\newcommand{\mPh}{ \mathbf{\Phi} }
\renewcommand{\l}{\ell}

\section{Introduction}
One of the major challenges in modern data analysis is to find low-dimensional structure in large high-dimensional data sets. A prevalent low-dimensional structure is that of data points lying in a union of subspaces. 
The problem of extracting such a structure from a given data set can be formalized as follows. 
Consider the (high-dimensional) set $\YS$ of points in $\reals^m$ and assume that 
$
\YS = \YS_1 \cup ...  \cup  \YS_L
$ 
where the points in $\YS_\l$ lie in a low-dimensional linear 
subspace $\cS_\l$ of $\reals^m$. 
The association of the data points to the $\YS_
\l$, and the orientations and dimensions of the subspaces $\cS_\l$ are all unknown. 
The problem of identifying the assignments of the points in $\YS$ to the $\YS_\l$ is referred to in the literature as subspace clustering \cite{vidal_subspace_2011} or hybrid linear modeling and has applications, inter alia, in unsupervised learning, image representation and segmentation, computer vision, and disease detection. 

In practice one may have access to dimensionality-reduced observations of $\YS$ only, resulting, e.g., from ``undersampling'' due to complexity and speed constraints on the acquisition device. 
More pertinently, even if the data points in $\YS$ are directly accessible, it is often desirable to perform clustering in a lower-dimensional space 
as this reduces data storage costs and leads to computational complexity savings.  
 The idea of reducing computational complexity through dimensionality reduction 
  appears, e.g., in \cite{vempala_random_2005} in a general context, and for subspace clustering in the experiments reported in \cite{zhang_hybrid_2012,elhamifar_sparse_2013}.  Dimensionality reduction also has a privacy-enhancing effect in the sense that no access to the original data is needed for processing \cite{liu_random_2006}. 

A widely used mathematical tool in the context of dimensionality reduction is the Johnson-Lindenstrauss Lemma \cite{johnson_extensions_1984}, which states that an $N$-point set in Euclidean space can be embedded via a suitable linear map into a $O(\epsilon^{-2} \log N)$-dimensional space while preserving the pairwise Euclidean distances between the points up to a factor of $1 \pm \epsilon$. 
Random projections satisfy the properties of this linear map with high probability, which explains the popularity of the so-called random projection method \cite{vempala_random_2005}. 

%
%
%
Dimensionality reduction will, in general, come at the cost of clustering performance. The purpose of the present paper is to analytically characterize this performance degradation for two subspace clustering algorithms, namely sparse subspace clustering (SSC) \cite{elhamifar_sparse_2009,elhamifar_sparse_2013} and thresholding based subspace clustering (TSC) \cite{heckel_robust_2013}. 
Both SSC and TSC were shown to provably succeed  under very general conditions on the high-dimensional data set to be clustered, in particular even when the subspaces $\cS_\l$ intersect. The corresponding analytical results in \cite{soltanolkotabi_geometric_2011,soltanolkotabi_robust_2013,heckel_robust_2013} form the basis for quantifying the impact of dimensionality reduction on clustering performance.

\paragraph*{Notation}
We use lowercase boldface letters to denote (column) vectors and uppercase boldface letters to designate matrices. 
The superscript $\herm{}$ stands for transposition. 
For the vector $\vx$, $x_q$ denotes its $q$th entry. 
For the matrix $\mA$, $\mA_{ij}$ designates the entry in its $i$th row and $j$th column, 
 $\norm[2\to 2]{\mA} \defeq\;$ $\max_{\norm[2]{\vv} = 1  } \norm[2]{\mA \vv}$ its spectral norm, and $\norm[F]{\mA} \defeq (\sum_{i,j} |\mA_{ij}|^2 )^{1/2}$ its Frobenius norm. 
The identity matrix is denoted by $\mI$. 
 $\log(\cdot)$ refers to the natural logarithm, $\acos(\cdot)$ is the inverse function of $\cos(\cdot)$, and $x \land y$ stands for the minimum of $x$ and $y$. 
The set $\{1,...,N\}$ is denoted by $[N]$, and the cardinality of the set $\S$ is written as $|\S|$.
 $\mathcal N( \boldsymbol{\mu},\boldsymbol{\Sigma})$ stands for the distribution of a Gaussian random vector with mean $\boldsymbol{\mu}$ and covariance matrix $\boldsymbol{\Sigma}$. 
The unit sphere in $\reals^m$ is $\US{m} \defeq \{ \vx \in \reals^m \colon \norm[2]{\vx} = 1 \}$. 

\section{Formal problem statement and contributions}

Consider a set of data points in $\reals^m$, denoted by $\YS$, and assume that 
$
\YS = \YS_1 \cup ...  \cup  \YS_L
$, 
where the points $\vy_i^{(\l)} \in \YS_\l, i\in [n_\l]$, lie in a $d_\l$-dimensional linear subspace of $\reals^m$, denoted by $\cS_\l$. 
We consider a semi-random data model with deterministic subspaces $\cS_\l$ and the data points $\vy_i^{(\l)}$ sampled uniformly at random from $\cS_\l \cap \US{d_\l}$. Neither the assignments of the points in $\YS$ to the sets $\YS_\l$ nor the subspaces $\cS_\l$ are known. 
%
%
Clustering of the points in $\YS$ is performed by first applying the (same) realization of a random matrix $\mPh \in \reals^{p\times m}$, $p\leq m$, typically $p  \ll m$, to each point in $\YS$ to obtain the set of dimensionality-reduced data points $\X$, and then declaring the segmentation obtained by SSC or TSC applied to $\X$ to be the segmentation of the data points in $\YS$. The realization of $\mPh$ does not need to be known. 
There are two error sources that determine the performance of this approach. 
First, the error that would be obtained even if clustering was performed on the high-dimensional data set $\YS$ directly. Second, and more pertinently, the error incurred by operating on dimensionality-reduced data. 
 The former is quantified for SSC in \cite{soltanolkotabi_geometric_2011,soltanolkotabi_robust_2013} and for TSC in \cite{heckel_robust_2013}, while characterizing the latter analytically is the main contribution of this paper. 
%
Specifically, we find that SSC and TSC applied to the dimensionality-reduced data set $\X$ provably succeed under quite general conditions on the relative orientations of the subspaces $\cS_\l$, provided that $\YS$ (and hence $\X$) contains sufficiently many points from each subspace. 
Our results make the impact of dimensionality-reduction explicit and show that SSC and TSC succeed even if $p$ is on the order of the dimensions of the subspaces. 
Moreover, we reveal a tradeoff between the affinity of the subspaces and the amount of dimensionality-reduction possible. 
The mathematical engine behind our theorems is a result stating that 
randomly projecting $d$-dimensional subspaces (of $\reals^m$) into $p$-dimensional space does not increase their affinities by more than const.$\sqrt{d/p}$, with high probability. 
Finally, we provide numerical results quantifying the impact of dimensionality reduction through random projection on algorithm running-time and clustering performance. 

%
%

\vspace{-0.1cm}
\section{SSC and TSC}

We next briefly summarize the SSC \cite{elhamifar_sparse_2009,elhamifar_sparse_2013} and TSC \cite{heckel_robust_2013} algorithms, both of which are based on 
 the principle of  applying spectral clustering \cite{luxburg_tutorial_2007} to an adjacency matrix $\mA$  constructed from the data points to be clustered. 
In SSC $\mA$ is obtained by finding a sparse representation of each data point in terms of all the other data points via $\ell_1$-minimization (or via Lasso \cite{soltanolkotabi_robust_2013}). TSC constructs $\mA$ from  the nearest neighbors of each data point in spherical distance. 


{\bf The SSC algorithm:} Given a set of $N$ data points $\X$ in $\reals^p$ and an estimate of the number of subspaces $\hat L$ (estimation of $L$ from $\X$ is discussed later), perform the following steps. 

{\bf Step 1: } Let $\mX \in \reals^{p \times N}$ be the matrix whose columns are the points in $\X$. For each $j \in [N]$ determine $\vz_j$ as a solution of
  \begin{align}
  \underset{\vz}{\text{minimize }} \norm[1]{\vz}  \text{ subject to }  \vx_j = \mX \vz \text{ and } z_j=0.
  \label{eq:minsscw}
  \end{align}
  Construct the adjacency matrix $\mA$ according to $\mA = \mZ + \herm{\mZ}$, where $\mZ = \mathrm{abs} ([\vz_1\,...\,\vz_N])$, and $\mathrm{abs}(\cdot)$ takes absolute values element-wise.  

{\bf Step 2:} Apply normalized spectral clustering \cite{ng_spectral_2001,luxburg_tutorial_2007} to $(\mA, \hat L)$. 
\vspace{0.11cm} 

{\bf The TSC algorithm:} 
Given a set of $N$ data points 
 $\X$ in $\reals^p$, an estimate of the number of subspaces $\hat L$ (again, estimation of $L$ from $\X$ is discussed later), and the parameter $q$ (the choice of $q$ is also discussed later), perform the following steps: 

{\bf Step 1:} For every $\vx_j \in \X$, find the set $\S_j \subset [N] \!\setminus\! j$ of cardinality $q$ defined through
\begin{equation*}
\left| \innerprod{\vx_j}{ \vx_i} \right| \geq \left| \innerprod{\vx_j}{ \vx_p} \right| \text{ for all }  i \in \S_j \text{ and all } p \notin \S_j
\end{equation*}
and let $\vz_j \in \reals^N$ be the vector with $i$th entry $\exp(-2\acos(\left| \innerprod{\vx_j }{ \vx_i }\right| / (\norm[2]{\vx_j} \norm[2]{\vx_i}) ))$ if $i\in \S_j$, and $0$ if $i\notin \S_j$. 
Construct the adjacency matrix $\mA$ according to $\mA = \mZ + \herm{\mZ}$, where $\mZ = [\vz_1\,...\,\vz_N]$. 

{\bf Step 2:} Apply normalized spectral clustering \cite{ng_spectral_2001,luxburg_tutorial_2007} to $(\mA, \hat L)$. 

Let the oracle segmentation of $\X$ be given by $\X = \X_1 \cup ...  \cup  \X_L$. 
  If each connected component \cite[Sec.~2.1]{luxburg_tutorial_2007} 
  in the graph $G$ with adjacency matrix $\mA$ corresponds exclusively to points from one of the sets $\X_\l$, spectral clustering will deliver the oracle segmentation \cite[Prop.~4; Sec.~7]{luxburg_tutorial_2007} and the clustering error (CE), i.e., the fraction of misclassified points,  will be zero. 
Since the CE is inherently hard to quantify, we will work with an intermediate, albeit sensible, performance measure, also used in \cite{heckel_robust_2013,soltanolkotabi_geometric_2011,soltanolkotabi_robust_2013}. Specifically, we  declare success if the graph $G$ (with  adjacency matrix $\mA$ 
 obtained by the corresponding clustering algorithm) has no false connections,  i.e., each $\vx_j \in \X_\l$ is connected to points in $\X_\l$ only, for all $\l$. 
Guaranteeing the absence of false connections, does, however, not guarantee 
 that the connected components correspond to the $\X_\l$, as the points in a given set $\X_\l$ may be split up into two (or more) distinct clusters. 
TSC counters this problem by imposing that each point in $\X_\l$ is connected to at least $q$ other points in $\X_\l$ (recall that $q$ is the input parameter of TSC). 
Increasing $q$ reduces the chance of clusters splitting up, but at the same time also increases the probability of false connections. 
A procedure for selecting $q$ in a data-driven fashion is described in \cite{heckel_neighborhood_2014}. 
For SSC, provided that $G$ has no false connections, 
by virtue of $\vx_i = \mX \vz_i$, we automatically get (for non-degenerate situations\footnote{
Non-degenerate simply means that $d_\l$ points are needed to represent $\vx_i \in \X_\l$ through points in $\X_\l \setminus \vx_i$. 
This condition is satisfied with probability one for the statistical data model used in this paper. 
}) that each node corresponding to a point in $\X_\l$ is connected to at least $d_\l$ other nodes corresponding to $\X_\l$. 
 
  

For both SSC and TSC, the number of subspaces $L$ can be estimated based on the insight that the number of zero eigenvalues of the normalized Laplacian of $G$ is equal to the number of connected components of $G$ \cite{spielman_spectral_2012}. A robust estimator for $L$ is the \emph{eigengap heuristic} \cite{luxburg_tutorial_2007}.  


\section{\label{sec:perfguar} Main results}

We start by specifying the statistical data model used throughout the paper. 
The subspaces $\cS_\l$ are taken to be deterministic and the points within the $\cS_\l$ are chosen randomly. Specifically, the elements of the set $\YS_\l$ in $\YS = \YS_1 \cup ...  \cup  \YS_L$ are 
 obtained by choosing $n_\l$ points at random according to $\vy_j^{(\l)} = \mU^{(\l)} \va^{(\l)}_j, j \in [n_\l]$, where $\mU^{(\l)} \in \reals^{m\times d_\l}$ is an orthonormal basis for the $d_\l$-dimensional subspace $\cS_\l$, and the $\va^{(\l)}_j$ are i.i.d.~uniform on $\US{d_\l}$. Since each $\mU^{(\l)}$ is orthonormal, the data points $\vy_j^{(\l)}$ are distributed uniformly  on the set $\{\vy \in \cS_\l \colon \norm[2]{\vy} = 1 \}$. 
The data set $\X$ in the lower-dimensional space $\reals^p$ is obtained by applying the (same) realization of a random matrix $\mPh \in \reals^{p\times m}$ to each point in $\YS$. 
The elements of the sets $\X_\l$ in $\X = \X_1 \cup ... \cup \X_L$ are hence given by $\vx_j^{(\l)} = \mathbf{\Phi} \vy_j^{(\l)}, j \in [n_\l]$. 

We take $\mPh$ as a random matrix satisfying, for all $t>0$,
\begin{align}
\PR{ \left| \norm[2]{\mPh \vx}^2 - \norm[2]{\vx}^2 \right|  \geq t \norm[2]{\vx}^2 } \leq 2 e^{- \tilde c t^2 p}, \quad \forall  \, \vx \in \reals^{m}
\label{eq:conceqcondonPh}
\end{align}
where $\tilde c$ is a constant. 
The Johnson-Lindenstrauss (JL) Lemma is a direct consequence of \eqref{eq:conceqcondonPh} (see e.g., \cite{vempala_random_2005}). A random matrix satisfying \eqref{eq:conceqcondonPh} is therefore said to exhibit the JL property, which holds, 
inter alia, for matrices with i.i.d.~subgaussian\footnote{A random variable $x$ is subgaussian \cite[Sec.~7.4]{foucart_mathematical_2013} if its tail probability satisfies $\PR{|x| >t} \leq c_1 e^{-c_2 t^2}$ for constants $c_1, c_2>0$. Gaussian and Bernoulli random variables are subgaussian. 
}
  entries 
  \cite[Lem.~9.8]{foucart_mathematical_2013}. 
  Such matrices may, however, be costly to generate, store, and apply to the high-dimensional data points. 
  In order to reduce these costs structured random matrices satisfying \eqref{eq:conceqcondonPh} (with $\tilde c$ mildly dependent on $m$) were proposed in \cite{ailon_almost_2013,krahmer_new_2011}. 
An example of such a structured random matrix \cite{ailon_almost_2013} is the product of a partial Hadamard matrix $\mH\in \reals^{p\times m}$, obtained by choosing a set of $p$ rows uniformly at random from a Hadamard matrix, and a diagonal matrix $\mD\in \reals^{m\times m}$ with main diagonal elements drawn i.i.d.~uniformly from $\{-1,1\}$. By \cite[Prop.~3.2]{krahmer_new_2011}, the resulting matrix $\mH \mD$  
 satisfies \eqref{eq:conceqcondonPh} with $\tilde c = c_2 \log^{-4}(m)$, where $c_2$ is a numerical constant. Moreover, $\mH \mD$ can be applied in time $O( m \log m)$ as opposed to time $O(mp)$ for a subgaussian random matrix. 
The fact that  $\mH\mD$ satisfies \eqref{eq:conceqcondonPh} relies on a connection between \eqref{eq:conceqcondonPh} and the restricted isometry property (RIP), widely used in compressed sensing \cite{candes_introduction_2008}. 
Specifically, it follows from \cite[Thm.~9.11]{foucart_mathematical_2013}, that \eqref{eq:conceqcondonPh} implies the RIP, while  
conversely, \cite[Prop.~3.2]{krahmer_new_2011} establishes that randomization of the column signs of a matrix satisfying the RIP yields a matrix satisfying \eqref{eq:conceqcondonPh}.

%

The performance guarantees we obtain below 
 are in terms of the affinity between the subspaces $\cS_k$ and $\cS_\l$ defined as 
\cite[Def.~2.6]{soltanolkotabi_geometric_2011}, \cite[Def.~1.2]{soltanolkotabi_robust_2013} 
$
\aff(\cS_k,\cS_\l) \defeq \frac{1}{\sqrt{d_k \land d_\l }}    \big\| \herm{\mU^{(k)}} \mU^{(\l)}  \big\|_F. 
$ 
Note that $0 \leq \aff(\cS_k,\cS_\l) \leq 1$, with $\aff(\cS_k,\cS_\l) = 1$ if $\cS_k \subseteq \cS_\l$ or $\cS_\l \subseteq \cS_k$ and $\aff(\cS_k,\cS_\l) = 0$ if $\cS_k$ and $\cS_\l$ are orthogonal to each other. 
Moreover, $\aff(\cS_k,\cS_\l) =\allowbreak \sqrt{ \cos^2( \theta_1) + ...+ \cos^2(\theta_{d_k \land d_\l})}/\sqrt{d_k \land d_\l}$, where $\theta_1 \leq ... \leq \theta_{d_k \land d_\l }$ are the principal angles between $\cS_k$ and $\cS_\l$. 
If $\cS_k$ and $\cS_\l$ intersect in $t$ dimensions, i.e., if $\cS_k \cap \cS_\l$ is $t$-dimensional, then $\cos(\theta_1)=...=\cos(\theta_{t})=1$ and hence $\aff(\cS_k,\cS_\l) \geq \sqrt{t/ (d_k \land d_\l) }$. 

We start with our main result for SSC. 

\begin{theorem}
\label{th:RPNoiseless}
Suppose that $\rho_\l \defeq (n_\l-1)/d_\l\geq \rho_0$, for all $\l$, where $\rho_0$ is a numerical constant, and pick any $\tau>0$. Set $d_{\max} = \max_\l d_\l$, $\rho_{\min} = \min_\l \rho_\l$, and suppose that 
\begin{align}
\label{eq:ThmAffConditionrp}
\max_{k,\l  \in [L ]\colon  k\neq \l} \!\!\!\!
\mathrm{aff}(\cS_k,\cS_\l) 
\!+\! \frac{\sqrt{28 d_{\max} \!+\! 8 \log L \!+\! 2\tau }}{\sqrt{3 \tilde cp}}
\leq \frac{ \sqrt{\log \rho_{\min}} }{65 \log N  } 
\end{align}
where $\tilde c$ is the constant in \eqref{eq:conceqcondonPh}. 
Then, the graph $G$ with adjacency matrix $\mA$ obtained by applying  SSC to $\X$ has no false connections with probability at least 
$1 - 4e^{-\tau/2} - N^{-1}  -  \sum_{\l=1}^L n_\l e^{-\sqrt{\rho_\l}  d_\l}$. 
\end{theorem}

Our main result for TSC is the following. 
\begin{theorem}
\label{thm:tscrp}
Choose $q$ such that $n_\l \geq 6 q$, for all $\l$. If 
\begin{align}
\max_{k,\l  \in [L ]\colon  k\neq \l}
\aff(\cS_k,\cS_\l)  
+ \frac{\sqrt{10}}{\sqrt{12 \tilde c}}  \frac{\sqrt{d_{\max}}}{\sqrt{p} } \leq \frac{1}{15 \log N }
\label{eq:adikqopol}
\end{align}
where $\tilde c$ is the constant in \eqref{eq:conceqcondonPh}.  
Then, the graph $G$ with adjacency matrix $\mA$ obtained by applying TSC to $\X$ has no false connections with probability at least $1 - 7 N^{-1} - \sum_{\l=1}^L n_\l e^{-c(n_\l-1)}$, where $c>1/20$ is a numerical constant. 
\end{theorem}

Proof sketches of Theorems \ref{th:RPNoiseless} and \ref{thm:tscrp} can be found in Sections \ref{sec:proofRPBP} and \ref{sec:proofRPTSC}, respectively. 
The mathematical engine behind Theorems \ref{th:RPNoiseless} and \ref{thm:tscrp} is a result stating that randomly projecting a pair of $d$-dimensional subspaces (of $\reals^m$) into $p$-dimensional space, using a projection matrix satisfying the JL property, does not increase their affinity by more than const.$\sqrt{d/p}$, with high probability. 
Theorems \ref{th:RPNoiseless} and \ref{thm:tscrp} essentially state that SSC and TSC succeed with high probability if the affinities between the subspaces $\cS_\l$ are sufficiently small, if $\YS$ (and hence $\X$) contains sufficiently many points from each subspace, and if $p$ is not too small relative to  $d_{\max}$. 
Specifically, $p$ may be taken to be linear (up to log-factors) in the dimensions of the subspaces $\cS_\l$. 
%
Comparing to the clustering conditions for SSC \cite[Thm.~2.8]{soltanolkotabi_geometric_2011} and TSC \cite[Thm.~2]{heckel_robust_2013} when applied to the original data set $\YS$, we conclude that the impact of dimensionality reduction through projections satisfying the JL property is essentially quantified by adding a term proportional to $\sqrt{d_{\max}/p}$ to the maximum affinity between the subspaces $\cS_\l$. 
%
Conditions \eqref{eq:ThmAffConditionrp} and \eqref{eq:adikqopol} hence nicely reflect the intuition that the smaller the affinities between the subspaces $\cS_\l$, the more aggressively we can reduce the dimensionality of the data set without compromising performance. 
%

\section{Proof Sketch of Theorem \ref{th:RPNoiseless}\label{sec:proofRPBP}}
\newcommand{\p}{p} 

The proof is based on the following generalization of a result by Soltanolkotabi and Cand\`es \cite[Thm.~2.8]{soltanolkotabi_geometric_2011} from orthonormal bases $\mV^{(\l)}$  to arbitrary bases $\mV^{(\l)}$  for $d_\l$-dimensional subspaces of $\reals^p$.

\begin{theorem}
\label{th:probrec}
Suppose that the elements of the sets $\X_\l$ in $\X = \X_1 \cup ...  \cup  \X_L$ 
are obtained by choosing $n_\l$ points at random according to $\vx_j^{(\l)} = \mV^{(\l)} \va^{(\l)}_j, j \in [n_\l]$, where the $\mV^{(\l)} \in \reals^{\p\times d_\l}$ are deterministic matrices of full rank and the $\va^{(\l)}_j$ are i.i.d.~uniform on $\US{d_\l}$. Assume that $\rho_\l \defeq (n_\l-1)/d_\l\geq \rho_0$, for all $\l$, where $\rho_0$ is a numerical constant, and let $\rho_{\min} = \min_\l \rho_\l$. If
\begin{equation}
\label{eq:ThmAffCondition2}
\max_{k,\l  \in [L ]\colon  k\neq \l}  \frac{1}{\sqrt{ d_k}  }   \norm[F]{\pinv{\mV^{(\l)}}  \mV^{(k)}}
  \leq \frac{ \sqrt{\log \rho_{\min}} }{ 64 \log N}
\end{equation}
where $\pinv{\mV^{(\l)}} = \inv{(\transp{\mV^{(\l)}} \mV^{(\l)}  )}\transp{\mV^{(\l)}}$ is the pseudo-inverse of $\mV^{(\l)}$, 
then the graph $G$ with adjacency matrix $\mA$ obtained by applying SSC to $\X$ has no false connections with probability at least 
$1 -  N^{-1}  -  \sum_{\l=1}^L n_\l e^{-\sqrt{\rho_\l}  d_\l}$. 
\end{theorem}

The proof of Theorem \ref{th:probrec}, not given here, essentially follows that of \cite[Thm.~2.8]{soltanolkotabi_geometric_2011} with minor changes. 

Set $\mV^{(\l)} = \mPh \mU^{(\l)}$ in Theorem \ref{th:probrec}. For $\Ph = \mI$ (which requires $p=m$) the LHS of \eqref{eq:ThmAffCondition2} reduces to $\max_{k,\l  \in [L ]\colon  k\neq \l} \aff(\cS_k,\cS_\l)$. 
Here, however, we need to work with the projected data, and $\mPh \mU^{(\l)}$ will in general not be orthonormal, which explains the need for the generalization to arbitrary bases $\mV^{(\l)}$. 
The columns of the matrix $\mV^{(\l)} \in \reals^{p\times m}$ ($\mV^{(\l)}$ has full column rank for $p \geq d_\l$ with high probability, not shown here) form a basis for the $d_\l$-dimensional subspace of $\reals^p$ containing the points in $\X_\l$. The proof of Theorem~\ref{th:RPNoiseless} is now effected by showing that randomly projecting the subspaces $\cS_k, \cS_\l \subseteq \reals^m$ into $p$-dimensional space 
through a matrix satisfying the JL property does not increase their affinity by more than const.$\sqrt{d_{\max}/p}$, with high probability. 
This can be formalized by first noting that
\begin{align}
&\frac{1}{\sqrt{d_k \land d_\l }} \norm[F]{\transp{\mV^{(\l)}}  \mV^{(k)}} \leq\nonumber \\
&\frac{1}{\sqrt{d_k \land d_\l }}
\left( \norm[F]{\transp{\mU^{(\l)}}  \mU^{(k)}} 
+
\norm[F]{\transp{\mV^{(\l)}} \mV^{(k)} - \transp{\mU^{(\l)}} \mU^{(k)} } \right) \nonumber \\
&=  \mathrm{aff}(\cS_k,\cS_\l) + \frac{1}{\sqrt{d_k \land d_\l }} \norm[F]{\transp{\mU^{(\l)}} ( \transp{\mPh} \mPh -\mI)  \mU^{(k)} } 
\label{eq:adsfasdfeioim}
\end{align}
and then showing that the ``perturbation'' 
$
 \frac{1}{\sqrt{d_k \land d_\l }} \norm[F]{\transp{\mU^{(\l)}} ( \transp{\mPh} \mPh -\mI)  \mU^{(k)} }
$ 
does not exceed const.$\sqrt{d_{\max}/p}$, with high probability. 
This result is then used to finalize the proof of Theorem \ref{th:RPNoiseless} 
by establishing that \eqref{eq:ThmAffConditionrp} implies \eqref{eq:ThmAffCondition2} with probability at least $1-4e^{-\tau}$. Set $\mQ_\l \defeq \inv{(\transp{\mV^{(\l)}} \mV^{(\l)})}$, for notational convenience, and note that the LHS of \eqref{eq:ThmAffCondition2} can be upper-bounded as follows 
\begin{align}
&\frac{1}{\sqrt{ d_k}} 
 \norm[F]{\pinv{\mV^{(\l)}}  \mV^{(k)}} 
 = \frac{1}{\sqrt{ d_k}  } \norm[F]{\mQ_\l \transp{\mV^{(\l)}}  \mV^{(k)}   } \nonumber \\
 &\leq \norm[2\to 2]{\mQ_\l}   \frac{1}{\sqrt{ d_k}  }   \norm[F]{\transp{\mV^{(\l)}}  \mV^{(k)}   } \nonumber \\
 &\leq \frac{\norm[2\to 2]{\mQ_\l}}{\sqrt{ d_k}  }\left( \norm[F]{\transp{\mU^{(\l)}}  \mU^{(k)}}   + \norm[F]{ \transp{\mU^{(\l)}} ( \transp{\mPh} \mPh -\mI)  \mU^{(k)}}   \right) \nonumber \\
  &\leq \norm[2\to 2]{\mQ_\l} \left( \mathrm{aff}(\cS_k,\cS_\l)  + \norm[2\to 2]{ \transp{\mU^{(\l)}} ( \transp{\mPh} \mPh -\mI)  \mU^{(k)}}   \right) \nonumber \\
&\leq \frac{1}{1-\delta} ( \mathrm{aff}(\cS_k,\cS_\l) + \delta )  \label{eq:withpradf} \\
&\leq \frac{65}{64} ( \mathrm{aff}(\cS_k,\cS_\l) + \delta ) \leq   \frac{ \sqrt{\log \rho_{\min}} }{64 \log N  } \label{eq:byassth:RPNoiseless}
\end{align}
where \eqref{eq:withpradf} holds with 
$
\delta \defeq \frac{\sqrt{28 d_{\max} + 8 \log L + 2\tau }}{\sqrt{3 \tilde cp}}
$
 with probability at least $1-4e^{-\tau}$ (not shown here), and for \eqref{eq:byassth:RPNoiseless} we used \eqref{eq:ThmAffConditionrp} twice (
note that since $\mathrm{aff}(\cS_k,\cS_\l) \geq 0$ and $\frac{ \sqrt{\log \rho_{\min}} }{\log N} \leq 1$, \eqref{eq:ThmAffConditionrp} implies $\delta \leq 1/65$, i.e., $\frac{1}{1-\delta} \leq \frac{65}{64}$).  
Note that  \eqref{eq:withpradf} is the formal version of \eqref{eq:adsfasdfeioim}. 
The probability estimates used to obtain \eqref{eq:withpradf} rely on \cite[Thm.~9.9, Rem.~9.10]{foucart_mathematical_2013}; for the special case of a Gaussian random matrix $\mPh$, these estimates can also be obtained using standard results on the extremal singular values of Gaussian random matrices.

\section{\label{sec:proofRPTSC}Proof Sketch of Theorem \ref{thm:tscrp} }

The proof follows closely that of Theorem~3 in \cite{heckel_robust_2013}. 
The graph $G$ with adjacency matrix $\mA$ obtained by applying TSC to $\X$ has no false connections, i.e., each $\vx_i^{(\l)}$ is connected to points in $\X_\l$ only,  if for each $\vx_i^{(\l)} \in \X_\l$ the associated set $\S_i$ corresponds to points in $\X_\l$ only, for all $\l$. This is the case if 
\begin{align}
z_{(n_\l - \q)}^{(\l)} > 
\max_{k \in [L]\setminus\l, j \in [n_k]}
z_{j}^{(k)}
\label{eq:tscsdpfox2}
\end{align}
where $z_{j}^{(k)} \defeq \big| \big< \vx_j^{(k)} ,  \vx_i^{(\l)} \big> \big|$, and $z_{(1)}^{(\l)} \leq z_{(2)}^{(\l)} \leq ...\leq z_{(n_\l-1)}^{(\l)}$ are the order statistics of $\{z_{j}^{(\l)}\}_{j \in [n_\l] \setminus i}$. 
Note that, for simplicity of exposition, the notation $z_j^{(k)}$ does not reflect dependence on $\vx_i^{(\l)}$. 
The proof is established by upper-bounding the probability of \eqref{eq:tscsdpfox2} being violated. A union bound over all $N$ vectors $\vx_i^{(\l)}, i\in [n_\l], \l\in [L]$, then yields the final result. 
We start by setting 
$
\tilde z_j^{(k)} \defeq  \left| \innerprod{  \vy_j^{(k)} }{ \vy_i^{(\l)}  } \right|
$, 
where $\vy_j^{(k)} = \mU^{(k)} \va_j^{(k)}$ are the data points in the high-dimensional space $\reals^m$, 
and noting that
$
z_j^{(k)} = \left| \innerprod{\vx_j^{(k)}}{ \vx_i^{(\l)} } \right| =   \left|  \innerprod{\vy_j^{(k)}}{ \vy_i^{(\l)} } +  e_j^{(k)}  \right|
$
with 
$
e_j^{(k)} \defeq  
\innerprod{ \Ph \vy_j^{(k)}}{ \Ph\vy_i^{(\l)} }   -  \innerprod{\vy_j^{(k)}}{ \vy_i^{(\l)} } 
= \innerprod{ (\transp{\Ph} \Ph - \mI )\vy_j^{(k)}}{\vy_i^{(\l)} }
=  \innerprod{ \transp{\mU^{(\l)}} (\transp{\Ph} \Ph - \mI )\mU^{(k)} \va_j^{(k)}}{\va_i^{(\l)} }
$. 
 The probability of \eqref{eq:tscsdpfox2} being violated can now be upper-bounded as
\begin{align}
&\PR{z_{(n_\l-\q)}^{(\l)} \leq \max_{k \in [L]\setminus\l, j \in [n_k]} 
z_{j}^{(k)} } 
\leq 
\PR{\tilde z_{(n_\l-\q)}^{(\l)} \leq \frac{2}{3\sqrt{d_\l}}   }  \nonumber \\
&+\PR{ \max_{k \in [L]\setminus\l, j \in [n_k]} 
\tilde z_{j}^{(k)}  \geq \alpha  }  
+  \!\!\sum_{(j,k)\neq (i,\l)} \!\! \PR{ \big|  e_j^{(k)} \big|   \,\geq \, \epsilon }  
\label{eq:labprostobubb}
\end{align}
where we assumed that $\alpha + 2 \epsilon \leq \frac{2}{3\sqrt{d_\l}}$, with 
$
\alpha \defeq \frac{\sqrt{6\log N}4\sqrt{\log N} }{\sqrt{d_\l}} \max_{k \in [L]\setminus \l} 
\frac{1}{\sqrt{d_k}}  \norm[F]{ \herm{\mU^{(k)}} \mU^{(\l)} }
$ 
and 
 $\epsilon \defeq \frac{\sqrt{6\log N}}{\sqrt{d_\l}} \delta$, where $\delta \defeq  \frac{\sqrt{28 d_{\max} + 8 \log L + 4 \log N }}{ \sqrt{ 3 \tilde c  p}}$. 
Resolving this assumption leads to  
\[
\max_{k \in [L]\setminus \l}
\frac{1}{\sqrt{d_\l}}  \norm[F]{ \herm{\mU^{(k)}} \mU^{(\l)} } + \frac{\delta}{4\sqrt{\log N}}  \leq \frac{2}{3\cdot 4 \sqrt{6}   \log N }
\]
which is implied by \eqref{eq:adikqopol}  (using that $\sqrt{28 d_{\max} + 8 \log L + 4 \log N }/\sqrt{\log N} \leq \sqrt{40 d_{\max}}$). 

We next show that the distortion $e_j^{(k)}$ caused by the random projection is small. Analogously to the proof of Theorem~\ref{th:RPNoiseless} this is accomplished by making use  
of the fact that the perturbation 
$
 \frac{1}{\sqrt{d_k \land d_\l }} \norm[F]{\transp{\mU^{(\l)}} ( \transp{\mPh} \mPh -\mI)  \mU^{(k)} }
$ 
does not exceed const.$\sqrt{d_{\max}/p}$. Specifically, note that 
\begin{align}
&\sum_{(j,k)\neq (i,\l)} \PR{ \big|  e_j^{(k)} \big| \geq \epsilon }  \nonumber \\
&\leq 
\PR{\max_{\l\neq k} \norm[2\to 2]{ \transp{\mU^{(\l)}} ( \transp{\mPh} \mPh -\mI)  \mU^{(k)}}  \geq \delta }  \nonumber \\
&+\hspace{-0.3cm}
\sum_{(j,k)\neq (i,\l)} \hspace{-0.3cm}\PR{ \big|  e_j^{(k)} \big| \geq \frac{\sqrt{6\log N}}{\sqrt{d_\l}} \norm[2]{ \transp{\mU^{(\l)}} (\transp{\Ph} \Ph - \mI )\mU^{(k)} \va_j^{(k)} }   } \nonumber \\
&\leq  2e^{- \tau/2} + N^2 2 e^{-\frac{6 \log N}{2}} = \frac{4}{N}
\label{eq:boundonerr}
\end{align}
where we used a probability estimate based on \cite[Thm.~9.9, Rem.~9.10]{foucart_mathematical_2013} and a standard concentration inequality (e.g.,~\cite[Ex.~5.25]{vershynin_introduction_2012}). 
Using standard concentration of measure results and the assumption $n_\l \geq 6  q$, the probabilities in \eqref{eq:labprostobubb} are upper-bounded according to Steps 1 and 2 in \cite[Proof of Thm.~3]{heckel_robust_2013} by $e^{-c (n_\l-1)}$ and $3N^{-2}$, respectively, where $c>1/20$ is a numerical constant. 
With \eqref{eq:labprostobubb} we thus get 
that \eqref{eq:tscsdpfox2} is violated with probability at most $e^{-c(n_\l-1)} +  7N^{-2}$. 
Taking the union bound over all vectors $\vx_i^{(\l)}, i \in [n_\l], \l \in [L]$, yields the desired lower bound on $G$ having no false connections.



\section{Numerical Results}

We evaluate the impact of dimensionality reduction on
the performance of SSC and TSC applied to the problem of clustering face
 images taken from the Extended Yale B data set \cite{georghiades_illumination_2001,lee_acquiring_2005}, 
which contains $192 \times 168$ pixel 
frontal face images of $38$ individuals, each acquired under $64$ different illumination conditions. 
The motivation for posing this problem as a subspace clustering problem comes from the insight that the vectorized images of a given face taken under varying illumination conditions lie near 9-dimensional linear subspaces \cite{basri_lambertian_2003}. 
In our terminology, each 9-dimensional subspace $\cS_\l$ would then correspond to an individual and would contain the images of that individual. 
For SSC, we use the implementation 
described in \cite{elhamifar_sparse_2013}, which is based on Lasso (instead of $\ell_1$-minimization) and uses the Alternating Direction Method of Multipliers (ADMM). Throughout this section, we set $q=4$ in TSC. 
Matlab code to reproduce the results below is available at 
http://www.nari.ee.ethz.ch/commth/research/. 

\begin{figure}
\begin{tikzpicture}[scale=1] 
    \begin{semilogxaxis}[ name=plot1,title = {running time},
    width =0.285\textwidth,
    legend entries={ {\scriptsize FRP}, {\scriptsize GRP}, {\scriptsize TSC}, {\scriptsize SSC}}, 
    legend style={
            cells={anchor=east},
                    legend pos= north west,}
    ]
    \addplot +[mark=none,black, dashed] table[x index=0,y index=1]{./times_faces_200it.dat};
    \addplot +[mark=none,black,dotted] table[x index=0,y index=2]{./times_faces_200it.dat};
    \addplot +[mark=none,black, dashdotted] table[x index=0,y index=3]{./times_faces_200it.dat};
    \addplot +[mark=none,black, solid] table[x index=0,y index=4]{./times_faces_200it.dat};
    \end{semilogxaxis}

    
    \begin{semilogxaxis}[name=plot2,at={($(plot1.east)+(1cm,0)$)},anchor=west, title = clustering error,
    width =0.285\textwidth,
    legend entries={ {\scriptsize GRP, TSC}, {\scriptsize GRP, SSC}, {\scriptsize FRP, TSC}, {\scriptsize FRP, SSC}}, 
    legend style={
            cells={anchor=east},
                    legend pos= north east,}
    ]
    \addplot +[mark=none,dashed, black] table[x index=0,y index=1]{./CERP_faces_200it.dat};
    \addplot +[mark=none,solid,black] table[x index=0,y index=2]{./CERP_faces_200it.dat};
    \addplot +[mark=none,dashdotted, black] table[x index=0,y index=3]{./CERP_faces_200it.dat};
    \addplot +[mark=none,dotted, black] table[x index=0,y index=4]{./CERP_faces_200it.dat};
    \end{semilogxaxis}

\end{tikzpicture}

\caption{\label{fig:fracesrprt} Running times and clustering error as a function of $p$.}

\vspace{-0.5cm}
\end{figure}
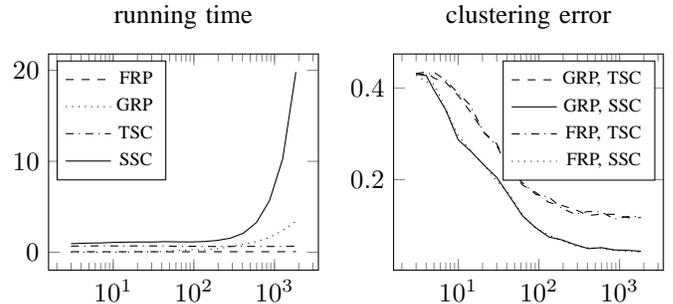

We generate $\YS$ by first selecting uniformly at random a subset of $\{1,...,38\}$ of cardinality $L=2$, and then collecting all images corresponding to the selected individuals. 
%
We use an i.i.d.~$\mathcal N(0,1/p)$  random projection matrix, referred to as GRP, 
and a fast random projection (FRP) matrix constructed similarly to the matrix $\mH \mD$ in Section \ref{sec:perfguar}. 
Specifically, we let $\mD \in \reals^{m\times m}$ be as in Section \ref{sec:perfguar} and 
take $\mF \in \complexset^{p\times m}$ to be a partial Fourier matrix obtained by choosing a set of $p$ rows uniformly at random from the rows of an $m \times m$ discrete Fourier transform (DFT) matrix. The FRP matrix is then given by the real part of $\mF \mD \in \complexset^{p\times m}$. 
In Figure \ref{fig:fracesrprt}, we plot the running times corresponding to the application of the GRP and the FRP matrix to the data set $\YS$, and the running times for TSC and SSC applied to the projected data $\X$, along with the corresponding CEs, as a function of $p$. 
For each $p$, the CE and the running times are obtained by averaging over $100$ problem instances (i.e., random subsets of $\{1,...,38\}$ and for each subset an independent realization of the random projection matrices). The results show, as predicted by Theorems \ref{th:RPNoiseless} and \ref{thm:tscrp},  that SSC and TSC, indeed, succeed provided that $d/p$ is sufficiently small (i.e., $p$ is sufficiently large). 
Moreover, SSC outperforms TSC, at the cost of larger running time. The running time of SSC increases significantly in $p$, while the running time of TSC does not increase notably in $p$. Since the FRP requires $O(m \log m)$ operations (per data point), its running time does not depend on $p$. Application of the GRP, in contrast, requires $O(mp)$ operations, and its running time exceeds that of TSC and SSC (applied to the projected data) for large $p$. 

\end{document}

%% file: rh_ss_def.tex
\usepackage{framed}

\newcommand\norm[2][\Tnorm]{\ensuremath{{\left\Vert #2 \right\Vert}_{#1}}}

\newcommand\Tinnerprod{}
\newcommand{\innerprod}[3][\Tinnerprod]{\ifthenelse{\equal{#1}{}}{\ensuremath{\left<#2,#3\right>}}{\ensuremath{\left<#2,#3\right>_{#1}}}}

\newcommand\vect[1]{\mathbf #1}

\definecolor{DarkBlue}{rgb}{0.1,0.1,0.5}
\definecolor{BrickRed}{RGB}{203,65,84}

\newcommand{\US}[1]{\mathbb S^{#1-1}} 

\newcommand\PR[1]{\ensuremath{ {\mathrm{P}}\!\left[#1\right]}}

\newcommand\Tex{}
\newcommand\EX[2][\Tex]{
\ifthenelse{\equal{#1}{}}{{\mathbb E}\left[#2\right]}{\ensuremath{{\mathbb E}_{#1}\left[ #2\right]}}}

\newcommand\Var[2][\Tex]{
\ifthenelse{\equal{#1}{}}{{\mathrm{Var} }[#2]}{\ensuremath{\mathrm{Var}_{#1}\left[ #2\right]}}}

\newcommand\ignore[1]{}

\newcommand\defeq{\coloneqq}

\newcommand{\reals}{\mathbb R} 
\newcommand{\complexset}{\mathbb C} 

\newcommand{\pinv}[1]{  {#1}^{ \dagger } } 
\newcommand{\inv}[1]{  {#1}^{ -1 } } 

\newcommand{\herm}[1]{{#1}^T} 
\newcommand{\transp}[1]{{#1}^T} 

\newcommand{\aff}{\mathrm{aff}}

\newcommand{\cS}{S} 
\newcommand{\X}{\mathcal X} 
\newcommand{\YS}{\mathcal Y} 

\newcommand{\va}{\vect{a}}

\newcommand{\vv}{\vect{v}}  

\newcommand{\vx}{\vect{x}}  
\newcommand{\vy}{\vect{y}}  
\newcommand{\vz}{\vect{z}}

\newcommand{\mA}{\vect{A}}

\newcommand{\mD}{\vect{D}}

\newcommand{\mF}{\vect{F}}

\newcommand{\mH}{\vect{H}}
\newcommand{\mI}{\vect{I}}

\newcommand{\mQ}{\vect{Q}}

\newcommand{\mU}{\vect{U}}
\newcommand{\mV}{\vect{V}}

\newcommand{\mX}{\vect{X}}

\newcommand{\mZ}{\vect{Z}}

\renewcommand{\S}{\mathcal T}

\newcommand{\q}{q} 

\renewcommand{\l}{{\ell}} 

\newcommand\acos{\mathrm{arccos}}

%% file: isit_2014_rev_v5.bbl
\begin{thebibliography}{10}
\providecommand{\url}[1]{#1}
\csname url@samestyle\endcsname
\providecommand{\newblock}{\relax}
\providecommand{\bibinfo}[2]{#2}
\providecommand{\BIBentrySTDinterwordspacing}{\spaceskip=0pt\relax}
\providecommand{\BIBentryALTinterwordstretchfactor}{4}
\providecommand{\BIBentryALTinterwordspacing}{\spaceskip=\fontdimen2\font plus
\BIBentryALTinterwordstretchfactor\fontdimen3\font minus
  \fontdimen4\font\relax}
\providecommand{\BIBforeignlanguage}[2]{{%
\expandafter\ifx\csname l@#1\endcsname\relax
\typeout{** WARNING: IEEEtran.bst: No hyphenation pattern has been}%
\typeout{** loaded for the language `#1'. Using the pattern for}%
\typeout{** the default language instead.}%
\else
\language=\csname l@#1\endcsname
\fi
#2}}
\providecommand{\BIBdecl}{\relax}
\BIBdecl

\bibitem{vidal_subspace_2011}
R.~Vidal, ``Subspace clustering,'' \emph{{IEEE} Signal Process. Mag.}, vol.~28,
  no.~2, pp. 52--68, 2011.

\bibitem{vempala_random_2005}
S.~S. Vempala, \emph{The Random Projection Method}.\hskip 1em plus 0.5em minus
  0.4em\relax American Mathematical Society, 2005.

\bibitem{zhang_hybrid_2012}
T.~Zhang, A.~Szlam, Y.~Wang, and G.~Lerman, ``Hybrid linear modeling via local
  best-fit flats,'' \emph{Int. J. Comput. Vision}, vol. 100, pp. 217--240,
  2012.

\bibitem{elhamifar_sparse_2013}
E.~Elhamifar and R.~Vidal, ``Sparse subspace clustering: Algorithm, theory, and
  applications,'' \emph{{IEEE} Trans. Pattern Anal. Machine Intell.}, vol.~35,
  no.~11, pp. 2765--2781, 2013.

\bibitem{liu_random_2006}
K.~Liu, H.~Kargupta, and J.~Ryan, ``Random projection-based multiplicative data
  perturbation for privacy preserving distributed data mining,'' \emph{{IEEE}
  Trans. Knowl. Data Eng.}, vol.~18, no.~1, pp. 92--106, 2006.

\bibitem{johnson_extensions_1984}
W.~B. Johnson and J.~Lindenstrauss, ``Extensions of {Lipschitz} mappings into a
  {Hilbert} space,'' \emph{Contemp. Math.}, no.~26, pp. 189--206, 1984.

\bibitem{elhamifar_sparse_2009}
E.~Elhamifar and R.~Vidal, ``Sparse subspace clustering,'' in \emph{Proc. of
  {IEEE} Conf. on Computer Vision and Pattern Recognition}, 2009, pp.
  2790--2797.

\bibitem{heckel_robust_2013}
R.~Heckel and H.~B\"olcskei, ``Robust subspace clustering via thresholding,''
  \emph{{arXiv:1307.4891}}, 2013, submitted to \emph{Ann. Stat.}

\bibitem{soltanolkotabi_geometric_2011}
M.~Soltanolkotabi and E.~J. Cand\`es, ``A geometric analysis of subspace
  clustering with outliers,'' \emph{Ann. Stat.}, vol.~40, no.~4, pp.
  2195--2238, 2012.

\bibitem{soltanolkotabi_robust_2013}
M.~Soltanolkotabi, E.~Elhamifar, and E.~J. Cand\`es, ``Robust subspace
  clustering,'' \emph{{arXiv:1301.2603}}, 2013, \emph{Ann.~Stat.}, accepted for
  publication.

\bibitem{luxburg_tutorial_2007}
U.~von Luxburg, ``A tutorial on spectral clustering,'' \emph{Stat. Comput.},
  vol.~17, no.~4, pp. 395--416, 2007.

\bibitem{ng_spectral_2001}
A.~Ng, I.~M. Jordan, and W.~Yair, ``On spectral clustering: Analysis and an
  algorithm,'' in \emph{Advances in Neural Information Processing Systems},
  2001, pp. 849--856.

\bibitem{heckel_neighborhood_2014}
R.~Heckel, E.~Agustsson, and H.~B\"olcskei, ``Neighborhood selection for
  thresholding based subspace clustering,'' in \emph{{Proc.~of} {IEEE}
  International Conference on Acoustics, Speech, and Signal Processing
  (ICASSP)}, 2014.

\bibitem{spielman_spectral_2012}
\BIBentryALTinterwordspacing
D.~Spielman, ``Spectral graph theory,'' 2012, lecture notes. [Online].
  Available: \url{http://www.cs.yale.edu/homes/spielman/561/}
\BIBentrySTDinterwordspacing

\bibitem{foucart_mathematical_2013}
S.~Foucart and H.~Rauhut, \emph{A mathematical introduction to compressive
  sensing}.\hskip 1em plus 0.5em minus 0.4em\relax Springer, Berlin,
  Heidelberg, 2013.

\bibitem{ailon_almost_2013}
N.~Ailon and E.~Liberty, ``An almost optimal unrestricted fast
  {Johnson-Lindenstrauss} transform,'' \emph{{ACM} Trans. Algorithms}, vol.~9,
  no.~3, pp. 1--12, 2013.

\bibitem{krahmer_new_2011}
F.~Krahmer and R.~Ward, ``New and improved {Johnson-Lindenstrauss} embeddings
  via the restricted isometry property,'' \emph{{SIAM} J. Math. Anal.},
  vol.~43, no.~3, pp. 1269--1281, 2011.

\bibitem{candes_introduction_2008}
E.~J.~Cand\`es and M.~Wakin, ``An introduction to compressive sampling,''
  \emph{{IEEE} Signal Proc. Mag.}, vol.~25, no.~2, pp. 21--30, Mar. 2008.

\bibitem{vershynin_introduction_2012}
R.~Vershynin, ``Introduction to the non-asymptotic analysis of random
  matrices,'' in \emph{Compressed sensing: Theory and applications}.\hskip 1em
  plus 0.5em minus 0.4em\relax Cambridge University Press, 2012, pp. 210--268.

\bibitem{georghiades_illumination_2001}
A.~S. Georghiades, P.~N. Belhumeur, and D.~J. Kriegman, ``From few to many:
  Illumination cone models for face recognition under variable lighting and
  pose,'' \emph{IEEE Trans. Pattern Anal. Mach. Intelligence}, vol.~23, no.~6,
  pp. 643--660, 2001.

\bibitem{lee_acquiring_2005}
K.~C. Lee, J.~Ho, and D.~J. Kriegman, ``Acquiring linear subspaces for face
  recognition under variable lighting,'' \emph{IEEE Trans. Pattern Anal. Mach.
  Intelligence}, vol.~27, no.~5, pp. 684--698, 2005.

\bibitem{basri_lambertian_2003}
R.~Basri and D.~Jacobs, ``Lambertian reflectance and linear subspaces,''
  \emph{{IEEE} Trans. Pattern Anal. Mach. Intell.}, vol.~25, no.~2, pp.
  218--233, 2003.

\end{thebibliography}
